\sfcode`A=1000 \sfcode`B=1000 \sfcode`C=1000 \sfcode`D=1000
\sfcode`E=1000 \sfcode`F=1000 \sfcode`G=1000 \sfcode`H=1000
\sfcode`I=1000 \sfcode`J=1000 \sfcode`K=1000 \sfcode`L=1000
\sfcode`M=1000 \sfcode`N=1000 \sfcode`O=1000 \sfcode`P=1000
\sfcode`Q=1000 \sfcode`R=1000 \sfcode`S=1000 \sfcode`T=1000
\sfcode`U=1000 \sfcode`V=1000 \sfcode`W=1000 \sfcode`X=1000
\sfcode`Y=1000 \sfcode`Z=1000
\documentstyle[12pt,aasms4] {article}

\begin{document}

\title{Time Evolution and the Nature of the Near-Infrared Jets in GRS1915+105}

\author{S.S. Eikenberry\altaffilmark{1} and G.G. Fazio\altaffilmark{1}}
\affil{Harvard-Smithsonian Center for Astrophysics, Cambridge, MA 02138}

\altaffiltext{1} {Visiting Astronomer, Kitt Peak National Observatory}

\begin{abstract}

	We observed the galactic microquasar GRS1915+105 in the K
($2.2 \mu$m) band on October 16 and 17, 1995 UTC using the COB
infrared (IR) imager on the Kitt Peak National Observatory 2.1-m
telescope with a 0.2-arcsec/pixel plate scale and under good ($\sim
0.7$-arcsec) seeing conditions.  Using a neighboring star in the image
frames to determine the point spread function (PSF), we PSF-subtract
the images of GRS1915+105.  We find no evidence of extended emission
such as the apparent near-IR jets seen by Sams {\it et al.} (1996) in
July, 1995.  Simple modelling of the star + jet structure allows us to
place an upper limit on any similar emission at that position of
$K>16.4$ at the 95\% confidence level, as compared to $K=13.9$ as seen
by Sams {\it et al.} (1996).  This lack of extended IR flux during
continued hard X-ray flaring activity confirms the hypothesis that the
extended IR emission arises from the superluminal radio-emitting jets
rather than reprocessing of the X-ray emission on other structures
around the compact central object.  Given the large apparent velocity
of the radio-emitting jets, by the time of our observations the Sams
{\it et al.} feature would have moved $>1$ arcsec from GRS1915+105,
and we can place a limit of $K>17.7$ (95\% confidence level) on any
infrared emission in this region.  We can thus place an upper limit on
the radiative timescale of the feature of $\tau < 25$ days, which is
consistent with synchrotron jet emission.

\end{abstract}

\keywords{infrared: stars -- stars: individual (GRS 1915+105) --ISM: jets and outflows}

\section{Introduction}

	The galactic microquasar GRS1915+105 is one of the most
intriguing objects in astrophysics today.  This object shows hard
X-ray, radio, and infrared (IR) flares with variability on timescales
from minutes to months (Mirabel and Rodriguez, 1996).  In addition,
VLA monitoring has revealed collimated ejection events exhibiting
superluminal motion interpreted as synchrotron emission from jets with
relativistic bulk motions (Mirabel and Rodriguez, 1994).  Recently,
Sams {\it et al.} (1996) have reported observations of extended
near-IR ($2.2 \mu$m) emission near the beginning of a hard X-ray flare
(Sazonov and Sunyaev, 1995).  While the position angle of the near-IR
extensions is consistent with the radio jets, the nature of the
emission is not well-determined.  Detailed multi-wavelength studies of
the flaring of GRS1915+105 imply the presence of other gas and/or dust
in the region surrounding the compact object, in addition to the jets
(Mirabel and Rodriguez, 1996).  Thus, possible sources of the extended
IR flux include reprocessing of the hard X-ray flare on an ejected gas
or dust disk, a wind outflow, or the radio jets, or synchrotron
emission from the jets themselves (Sams {\it et al.}, 1996).

\section{Observations}

	We observed GRS1915+105 with the COB infrared imager on the
Kitt Peak National Observatory\footnote{KPNO is operated by AURA Inc.
under contract to the National Science Foundation.} 2.1-m telescope on
October 16 and 17, 1995, roughly 3 months after Sams {\it et al.}
(1996).  We used the K-band ($2.2 \mu$m) filter, with a
0.2-arcsec/pixel plate scale.  On October 16, we took four 30-second
exposures through light cirrus, with seeing of $\sim 0.7$-arcsec
full-width half-maximum (FWHM).  On October 17, we again took four
30-second exposures, this time under photometric conditions, but with
$\sim 1.1$-arcsec FWHM seeing.  During each sequence of exposures, we
moved the telescope around the position of GRS1915+105 in order to
avoid defects in the array.  We subtracted the sky background and
flatfielded each exposure, and then combined each set of four
exposures.  Figure 1 shows a contour map of the resulting image for
the October 16 data.  Photometry of the combined images gives K-band
magnitudes of $K=13.51 \pm 0.05$ on October 16 and $K=13.43 \pm 0.03$
on October 17 (see also Eikenberry and Fazio, 1995).

\section{Analysis}

\subsection{Limits on point-like emission}

	In our analysis, we first concentrate on possible emission
from a feature similar to that seen by Sams {\it et al.} (1996).
While the separation between the stellar counterpart of GRS1915+105
and the extension of Sams {\it et al.} (1996) is less than 1/2 of our
image FWHM, if the feature's brightness ($K=13.9$) remained constant
it would be the source of 60\% of the photons in our images, and
simple PSF-subtraction should reveal its presence.  From each of the
background-subtracted and flatfielded (but uncombined) 30-second
exposures, we extract a $5 \times 5$-arcsec region centered on
GRS1915+105.  For our PSF, we extract an identical region centered on
a star near GRS1915+105 with a very similar K-band flux (Star A in
Figure 1).  For each exposure, we scale the PSF and subtract it from
the images of GRS1915+105.  We see no evidence for extended structure
in any of the PSF-subtracted exposures.  We then combine the four
PSF-subtracted exposures for each night.  Again, we find no evidence
for extended structure in the combined PSF-subtracted images (see
Figure 2).

	We then apply a simple sliding-cell source-detection algorithm
to the PSF-subtracted images.  In this approach, we take a model PSF
(a 2-D gaussian fit to Star A) and center it on a pixel in the image.
We multiply the image pixel values by the corresponding PSF values at
their location, and sum to obtain the total of the products.  We then
perform the same operation on the error image for the PSF-subtracted
image (including the uncertainties in the PSF), summing the error
products in quadrature.  The ratio of the image sum to the error sum
then gives the statistical significance of any point source at the
image pixel location.  Applying this algorithm to the PSF-subtracted
images from both nights, we find no source with a statistical
significance $>1 \sigma$ at any location.

	In order to estimate the sensitivity of our observations and
analysis techniques, we perform a simple Monte Carlo simulation.
First, we model the GRS1915+105 region as 2 point sources separated by
0.3-arcsec - the star plus a southern jet - as seen by Sams {\it et
al.} (1996).  We ignore the northern jet due to its much lower flux.
Second, (using the model PSF for both point sources), we select a
relative normalization of the model PSFs for the two point sources,
scale and add them with the appropriate positional offsets, and then
rescale the sum to give the same number of counts as in the real
GRS1915+105 image.  Next, we add normally-distributed random numbers
(having standard deviations determined from the quadrature-summed
uncertainties of both the GRS1915+105 and Star A (PSF) images) to each
pixel of the simulated image.  Finally, we take the image of Star A,
scale it, and subtract it from the simulated image, exactly as with
the actual GRS1915+105 images.

	In Figure 3, we present a typical simulated result of the PSF
subtraction for an extended component with $K=13.9$.  If the extended
emission seen by Sams {\it et al.} (1996) had remained unchanged, we
would have unambiguously found it in our data on both nights.  In
order to place an upper limit on any point-like flux at this position,
we then decrease the flux of the extended emission in the model and
repeat the simulation process.  We set our upper limit on the flux of
the extended component at the point where, for 100 Monte Carlo
simulations as described above, we detect the extended component in
the PSF-subtracted image at the 95\% confidence level using the
sliding-cell source-detection algorithm.  For October 16, the limit is
$K>16.4$, while for October 17 (when the seeing was poorer), the limit
is $K>15.8$.

\subsection{Limits on extended emission}

	The limit on point-like emission is useful in confirming that
the Sams {\it et al.} (1996) feature was transient, as expected for
emission from the radio-emitting jets.  However, since this feature is
associated with the radio-emitting jets of GRS1915+105, it will
exhibit superluminal motion, and may also expand at high velocities.
Thus, we have performed further analyses searching for possible
non-point-like emission from the jet.

	We perform this search using, once again, the sliding-cell
algorithm described above.  However, instead of using Star A as a PSF,
we use a broadened PSF for the jet emissions.  Applying the algorithm
to the stellar-PSF-subtracted images, we find no evidence of extended
emission using trial jet-PSFs with FWHM values of 0.8, 1.0, 1.25, and
1.5 arcsec.  Given that the Sams {\it et al.} feature was point-like
with their $\sim 0.2$ arcsec resolution, even if the feature expanded
at $0.5 c$ (much faster than the limt for the radio-emitting jets), it
could have expanded only to a FWHM of 1.25 arcsec, given the 12.5 kpc
distance to GRS1915+105 (Rodriguez {\it et al.}, 1995).  Thus, we
conclude that their is no evidence for infrared jet emissions in our
data.

	As with the point-like emission, we perform a Monte Carlo
simulation to estimate the sensitivity of our observations and
analysis techniques.  We now assume that the jet has moved 1.2 arcsec
farther from GRS1915+105 - an apparent velocity of $1.0c$, which is
less than is observed in the radio jets (Mirabel and Rodriguez, 1994).
We also assume a worst-case jet FWHM of 1.5 arcsec for the
source-detection algorithm - an expansion velocity $>0.5c$, which is
much greater than observed in the radio jets.  This approach gives an
upper limit of $K>17.7$ at the 95\% confidence level for any infrared
jet emission.

\section{Discussion}

	Given that the hard X-ray flaring activity which began in late
June or early July 1995 (Sazonov and Sunyaev, 1995) continued through
the time of these observations (Harmon {\it et al.}, 1995), the drop
of a factor $>10$ in the $2.2 \mu$m flux at the observed location of
Sams {\it et al.} (1996) places strong constraints on several of the
proposed explanations for the extended near-IR emission.  In
particular, hypotheses involving the reprocessing of the X-ray
emission on stellar winds, ejected dust or gas disks, or other
steady-state or slow-moving structures do not appear to explain such
behavior.  This, in addition to the appearance of the IR features
oppositely oriented about GRS1915+105, their position angle match with
the radio jets, and the similarities of the North/South flux asymmetry
to that in the radio (Sams et al., 1996), seems to confirm the
identification of the features seen by Sams {\it et al.} as infrared
jets.

	If the features are indeed due to infrared jets, then by the
time of our observations, the southern (bright) jet would have moved
more than 1 arcsec from GRS1915+105, and we have an upper limit of
$K>17.7$ in this region.  If the near-IR flux arises from reprocessing
of the X-ray emission on the jet, the reprocessing efficiency may have
dropped by this factor $>33$ due to the increased distance between the
X-ray source and the jet and/or changes in the X-ray opacity of the
jets.  Alternatively, if the IR flux arises from synchrotron processes
in the jet, then we can place an upper limit on the radiative lifetime
of the IR-emitting particles, using the time separation between our
observations and those of Sams {\it et al.} (1996) and the fact that
our upper limit is a factor 33 lower in flux than the Sams {\it et
al.} feature.  Thus, we find that the $1/e$ radiative lifetime of the
IR-emitting electroncs is $\tau <26$ days.  For synchrotron emission,
the relativistic electrons producing the IR emission have shorter
radiative lifetimes than the radio-emitting electrons by a factor of
$\sqrt{\nu_{IR} / \nu_{radio}} \sim 10^2$, independent of the magnetic
field strength.  Since the radio-emitting jets have lifetimes
significantly less than 1 year, we find that our limits are compatible
with the hypothesis that the Sams {\it et al.} feature arises from
synchrotron processes in the radio-emitting jets.

\section{Conclusions}

	We have presented near-infrared K ($2.2 \mu$m) band
observations of the galactic microquasar GRS1915+105 on October 16 and
17, 1995 with a 0.2 arcsec/pixel plate scale under good seeing
conditions.  Using PSF subtraction of the stellar image of
GRS1915+105, we find no evidence of near-infrared emissions as seen by
Sams {\it et al.} (1996) in July, 1995.  Simple modelling shows that
we would have detected any such extended emission at the 95\%
confidence level down to a limit of $K>16.4$, as compared to the
$K=13.9$ jet observed by Sams {\it et al.} (1996).  The fact that the
IR flux at this location dropped by a factor $>10$ during a time when
the hard X-ray flux increased seems to rule out reprocessing of the
hard X-ray emission on slow-moving or steady-state structures near the
compact object as a viable explanation for the extended IR emission,
and confirms the hypothesis that the extended IR emission arises from
the radio-emitting jets.

	If the features are indeed due to infrared jets, then by the
time of our observations, the southern (bright) jet would have moved
more than 1 arcsec from GRS1915+105, and we have an upper limit of
$K>17.7$ in this region, a factor of $>33$ drop in the IR flux.  This
allows us to place an upper limit on the radiative lifetime of the
feature of $\tau <26$ days.  These limits are consistent with the
hypothesis that the Sams {\it et al.} feature was due to synchrotron
processes in the radio-emitting jets of GRS1915+105.

\acknowledgements

	We would like to thank I.F. Mirabel for bringing the near-IR
jets in GRS1915+105 to our attention, M. Merrill for assisting with
the COB observations, and the anonymous referee for his/her helpful
comments.  S. Eikenberry is supported by a NASA Graduate Student
Researchers Program fellowship through Ames Research Center.

\begin{figure} 
\vspace*{200mm}
\includegraphics{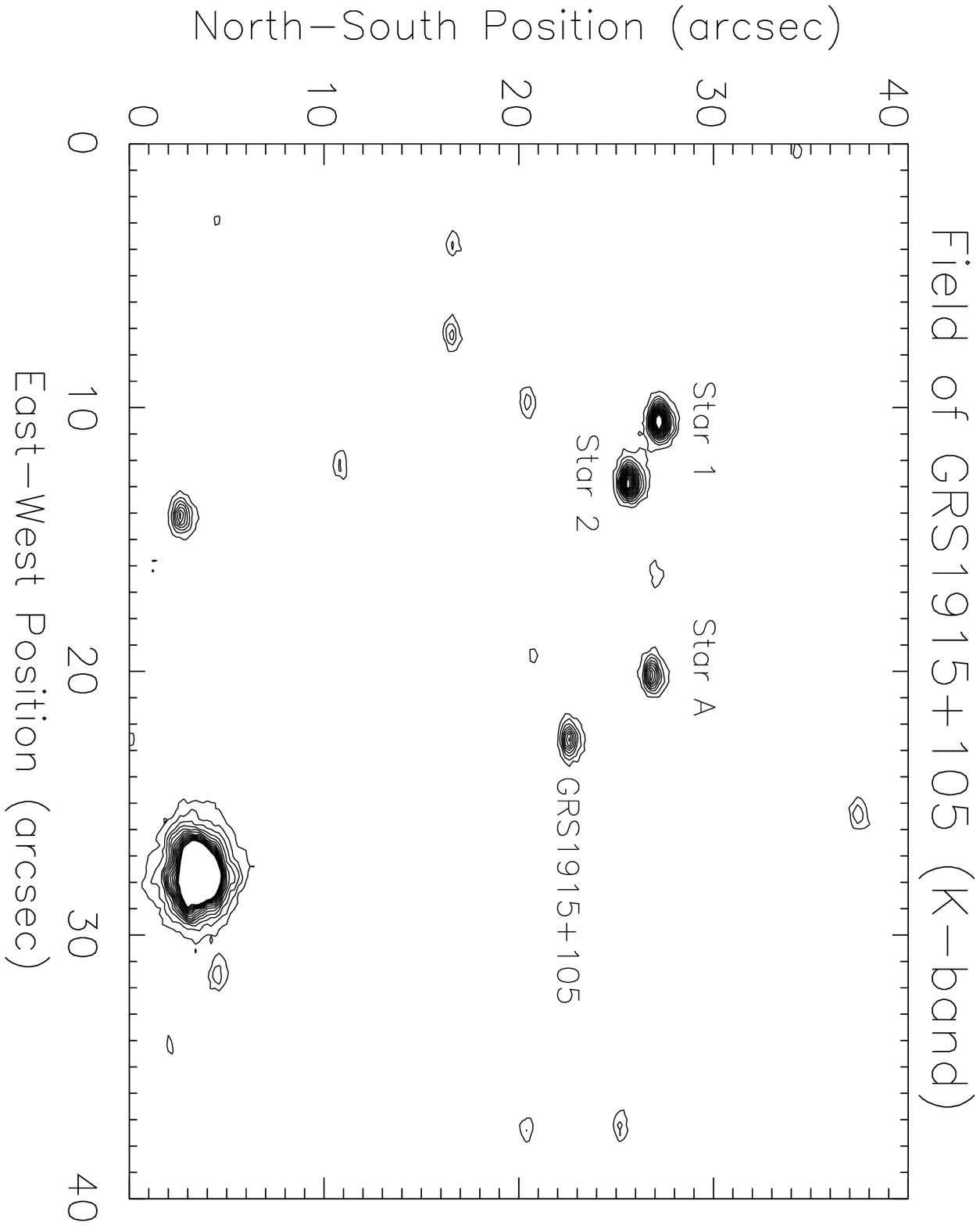} 
\caption{K-band ($2.2 \mu$m) map of the 40-arcsec field surrounding GRS1915+105} 
\end{figure} 

\begin{figure} \vspace*{200mm}
\includegraphics{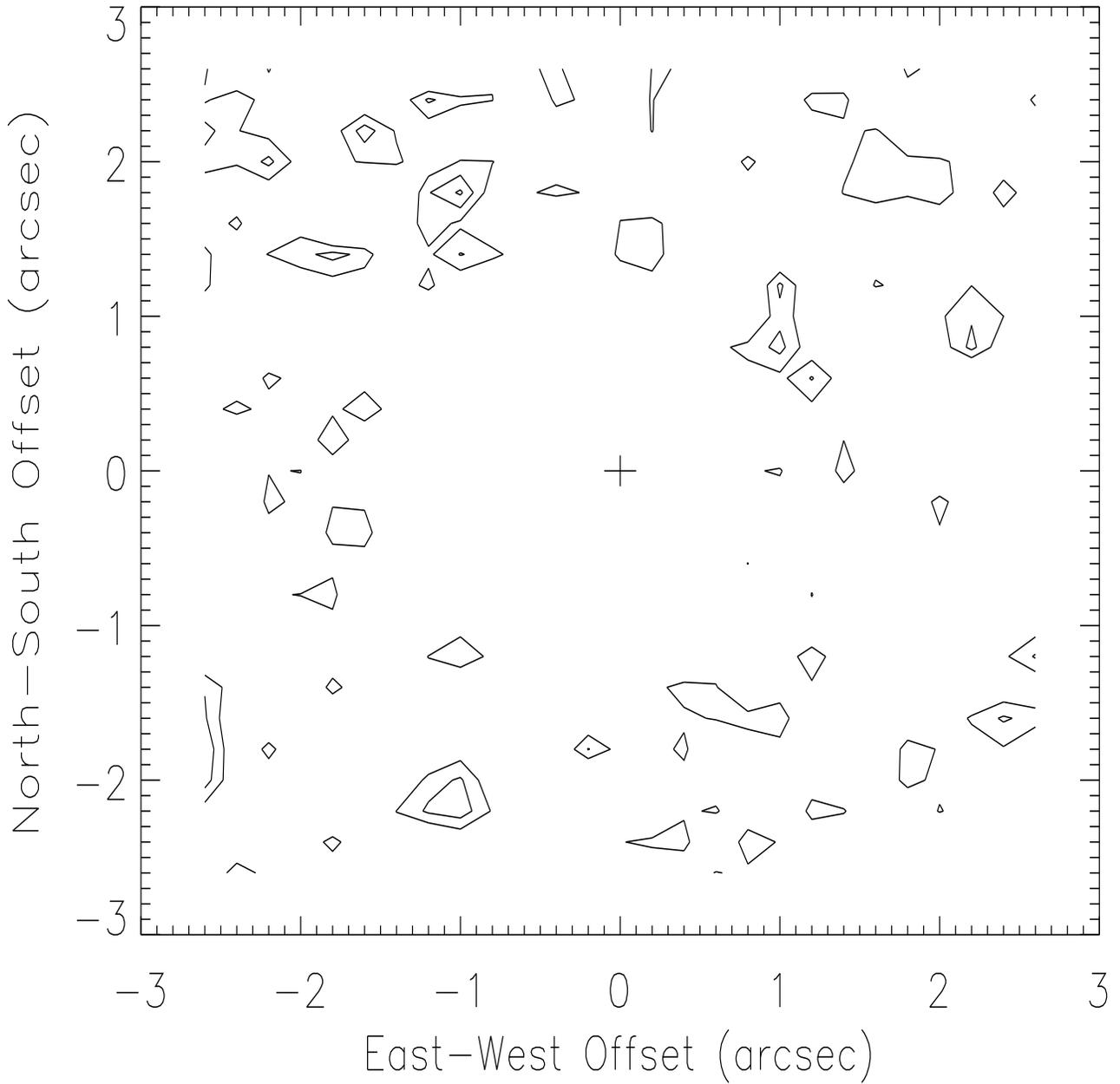} 
\caption{PSF-subtracted map of the 5-arcsec square region centered on GRS1915+105.  Contours are drawn at levels of $1\sigma, 2\sigma, 3\sigma, ...$ .  No significant residuals are evident.}  
\end{figure} 

\begin{figure} \vspace*{200mm}
\includegraphics{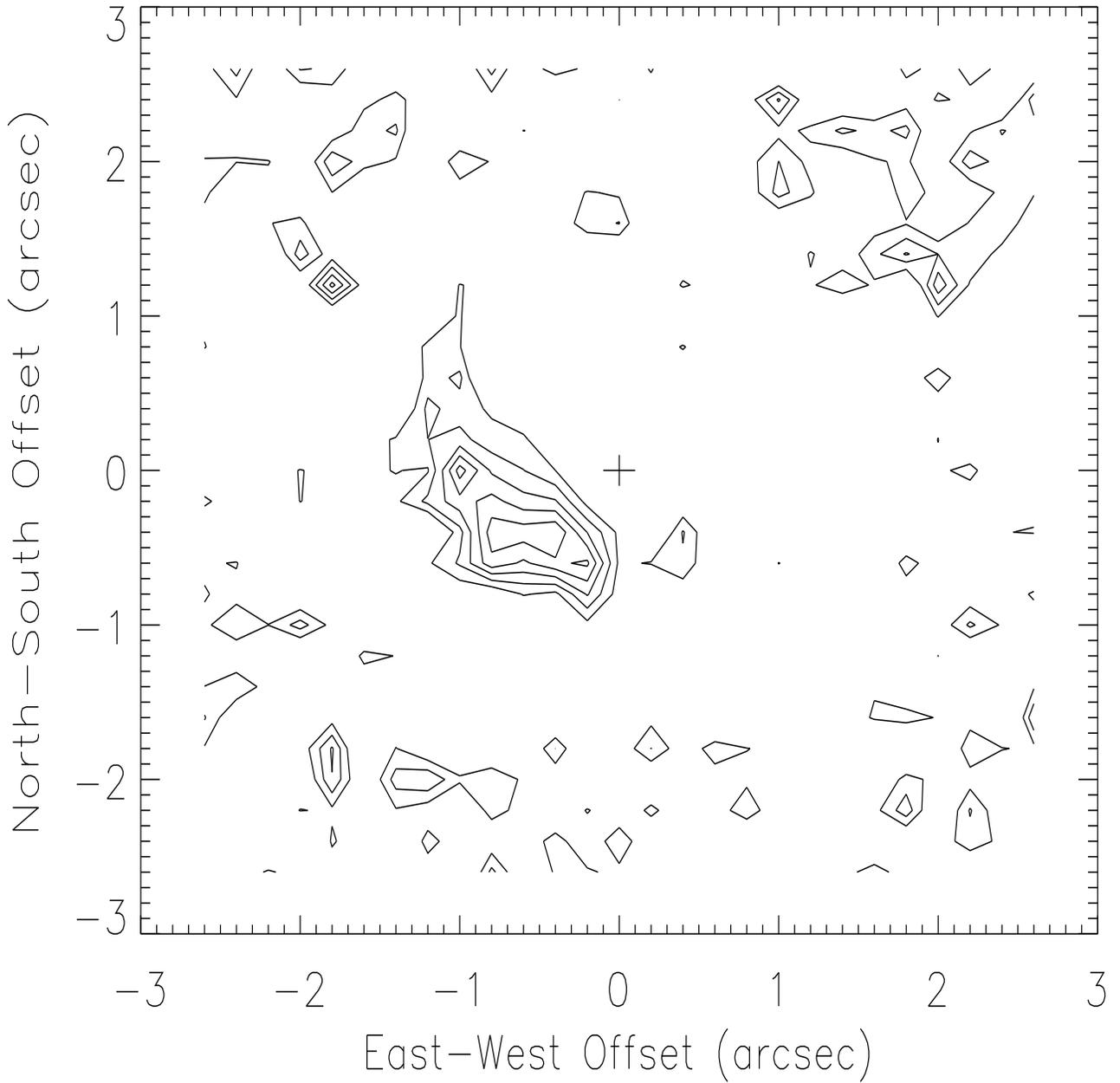} 
\caption{PSF-subtracted map of the 5-arcsec square region centered on a simulated image of GRS1915+105 including a jet component with K=13.9 at a 0.3-arcsec separation.  Contours are the same as in Figure 2.  The presence of such a jet component would be clearly evident if it were present in the actual data (Figure 2).}
\end{figure} 

\end{document}